\documentstyle[12pt]{article}\textheight 220mm\textwidth 150mm
\newcommand{\bi}{\bibitem}
\newcommand{\be}{\begin{eqnarray}}
\newcommand{\ee}{\end{eqnarray}}

\begin{document}

\catcode`\@=11
\def\lsim{\mathrel{\mathpalette\@versim<}}
\def\gsim{\mathrel{\mathpalette\@versim>}}
\def\@versim#1#2{\vcenter{\offinterlineskip
\ialign{$\m@th#1\hfil##\hfil$\crcr#2\crcr\sim\crcr } }}
\catcode`\@=12

\noindent
\hspace*{11.6cm}\vspace{-2mm}KANAZAWA-95-16\\
\hspace*{11.6cm}October 1995

\begin{center}
{\Large\bf  Gauge-Yukawa Unification\\
 and  \\
 the Top-Bottom  Hierarchy $ ^{\dag}$}
\end{center}

\vspace{0.3cm}

\begin{center}{\sc J. Kubo}$\ ^{(1)}$,
{\sc M.
Mondrag{\' o}n}$\ ^{(2)}$, \vspace{-1mm}\\
{\sc M. Olechowski}$\ ^{(3)}$ and
{\sc G. Zoupanos}$\ ^{(4),*}$
\end{center}
\begin{center}
{\em $\ ^{(1)}$
 College of Liberal Arts, Kanazawa \vspace{-2mm} University,
Kanazawa 920-11, Japan } \\
{\em $\ ^{(2)}$ Institut f{\" u}r Theoretische Physik,
Philosophenweg 16 \vspace{-2mm}\\
D-69120 Heidelberg, Germany}\\
{\em $\ ^{(3)}$ Institute of \vspace{-2mm} Theoretical Physics,
Warsaw University\\
ul. Hoza 69, 00-681 Warsaw, Poland}\\
{\em $\ ^{(4)}$ Physics Department, National
Technical\vspace{-2mm} University\\ GR-157 80 Zografou, Athens,
Greece }  \end{center}

\begin{center}
{\sc\large Abstract}
\end{center}

\noindent
The consequences of Gauge-Yukawa Unification (GYU)
in supersymmetric unified models
on low energy physics are analyzed.
We find that the observed top-bottom mass hierarchy
can be  explained
by supersymmetric GYU and  different
 models can be  experimentally distinguished if the top quark mass
lies slightly below its infrared value.

\vspace*{0.5cm}
\footnoterule
\vspace*{5mm}
\noindent
$ ^{\dag}$ Presented by G. Zoupanos at
{\em  Int. Europhysics Conf. on HEP},
Brussels, 27/7-2/8, 1995,
to appear in the proceedings.\\
$^{*}$Partially supported by the C.E.U. projects
(SC1-CT91-0729; CHRX-CT93-0319).

\newpage
\pagestyle{plain}
In Grand Unified Theories (GUTs),
 the gauge interactions of the standard model (SM) are unified at a
certain energy scale $M_{\rm GUT}$, and
this unification scheme
has given specific testable predictions \cite{gut1}.
The accurate
measurements of the gauge couplings at LEP in fact suggest that the
minimal $N=1$ supersymmetric $SU(5)$ GUT \cite{gut2} is
promising
 when comparing its theoretical values with
the experiments.

By a Gauge-Yukawa Unification (GYU) we mean a functional
relationship among the gauge and Yukawa couplings, which
can be derived from some principle.
In superstring and composite models for instance,
such relations
could be derived in principle.
In the GYU scheme \cite{mondragon1,kubo1,kubo2}, which is based on the
principle of finiteness and reduction of couplings,
one can write down
relations among the gauge and Yukawa couplings
 in a more concrete fashion.
(Note that the gauge and Yukawa sectors in
GUTs are usually not related.)
These principles are formulated
within the framework of
perturbatively renormalizable field theory, and
one can reduce the number of  independent
couplings without introducing necessarily a symmetry,
thereby improving the
calculability and predictive power of
a given theory\footnote{
In ref. \cite{jack}, interesting
renormalization group (RG) invariant relations among
the soft supersymmetry breaking parameters has been found.
These relations are obtained on the
close analogy of our approach presented here.}.

The consequence of GYU is that
in the lowest order in perturbation theory
 the gauge and Yukawa couplings above  $M_{\rm GUT}$
are related  in the form
\be
g_i& = &\kappa_i \,g_{\rm GUT}~,~i=1,2,3,e,\cdots,\tau,b,t~,
\ee
where $g_i~(i=1,\cdots,t)$ stand for the gauge
and Yukawa couplings, $g_{\rm GUT}$ is the unified coupling,
and
we have neglected  the Cabibbo-Kobayashi-Maskawa mixing
of the quarks.
 So, Eq. (1) exhibits a boundary condition on the
the renormalization group evolution for the effective theory
below $M_{\rm GUT}$, which we assume to
be the minimal supersymmetric standard model (MSSM).
 It has been recently found
\cite{mondragon1,kubo2} that various
supersymmetric GUTs with GYU in the
third generation can predict the bottom and top
quark masses that are consistent with the experimental data.
This means that the top-bottom hierarchy
could be
explained in these models,
exactly in the same way as
the hierarchy of the gauge couplings of the SM
can be explained if one assumes  the existence of a unifying
gauge symmetry at $M_{\rm GUT}$ \cite{gut1}.

It has been also observed \cite{mondragon1,kubo2}
that there exists a
relatively wide range of $k$'s which gives the
top-bottom hierarchy of the right order.
Of course, the existence of this range is partially related to the
infrared behavior of the Yukawa couplings \cite{hill1}.
 Therefore, a systematic investigation
on the nature of GYU is
indispensable
to see whether a
GYU can make experimentally distinguishable predictions on
the top and bottom masses, or whether the
top-bottom hierarchy results mainly from the infrared behavior of the
Yukawa couplings.
With more precise measurements of the top and bottom masses,
we will be able to  conclude which case is indeed realized.

We have
performed an exhaustive analysis on this problem
at the two-loop level \cite{kubo3},
and here we would like to
 present only a few representative results to provide an idea of our
complete analysis.
We have assumed that below $M_{\rm GUT}$ the evolution of couplings is
 governed by the MSSM and that there exists a unique threshold
$M_{\rm SUSY}$ for all superpartners of the MSSM so that
below $M_{\rm SUSY}$ the SM is the correct effective
theory, where we include only
the logarithmic and two-loop corrections
for the RG evolution of
couplings  \footnote{ When the threshold effects are
appropriately taken
into account,
the minimal supersymmetric
model based on $SU(5)$ predicts a value for the QCD
coupling at $M_Z$ that is slightly larger than
the experimental one \cite{threshold}.
Similar problem could exist here too. But we
ignore this problem, because
we do not consider any specific models.}.
We have neglected all the
threshold effects. Note that with a GYU boundary condition
alone the value of $\tan\beta$ can not be determined;
usually, it is determined in the Higgs sector, which however
strongly depends on the supersymmetry braking terms.
In our analysis we avoid this by using the tau
mass, along
with  $M_Z$, $\alpha_{\rm em}^{-1}(M_{Z})$ and $
\sin^{2} \theta_{\rm W}(M_{Z})$, as the input.

In Table, we present the predictions for
the $SU(5)$ type GYU
 (i.e. $  k_1=k_2=k_3=1,
k_b=k_{\tau}$) with $M_{\rm SUSY}=500$ GeV and $
k_b=\sqrt{(6/5)}\simeq 1.10$ fixed, where we vary $k_t$ from
$0.4$ to $2.0$. The Finite Unified Theory based on $SU(5)$
\cite{mondragon1}
corresponds to $k_t=\sqrt{(8/5)}\simeq 1.26$.

\begin{center}
{\bf Table }
\end{center}

\vspace{0.3cm}
\begin{tabular}{c|c|c|c|c}
\hline\hline
$k$ &$\alpha_{3}(M_Z)$ & $m_b (M_{b}) $ [GeV]
& $M_{t}$ [GeV] & $\tan\beta$
\\ \hline
$0.6$ & $0.118$ & $4.66$ & $150.2$ & $53.6$ \\  \hline
$0.8$ & $0.120$ & $4.65$ & $166.6$  &$53.5$ \\  \hline
$1.0$ & $0.121$ & $ 4.62 $ & $176.4$  & $53.6$ \\  \hline
$\sqrt{(8/5)}$ & $0.122$ & $4.57$ & $184.0$ & $53.8$ \\  \hline
$1.4$ & $0.122$ & $4.55$ & $186.5$  & $53.9$ \\  \hline
$1.6$ & $0.122$ & $4.51$ & $189.3$  & $54.1$ \\  \hline
$2.0$ & $0.123$ & $4.44$ & $192.9$   & $54.5$ \\  \hline\hline
\end{tabular}

\vspace{0.3cm}
$M_{t,b}$ are the pole masses while $m_b (M_b)$
is the running bottom quark
mass at its pole mass.
The values for $m_b (M_b)$ in the table should not be
taken very seriously because due to large values
of $\tan\beta$
there will be a relatively large
correction coming from the superpartner contribution
which is not included above.
We find that, because of the
infrared behavior of the Yukawa couplings \cite{hill1},
above $k_t\simeq 1.6$ (in the case at hand)  the
value of $M_t$ becomes  no longer sensitive  against the change
of $k_t$.
So, in general, if the experimental value for $M_t$ is
close to the infrared value,
 it is unclear whether the
top-bottom hierarchy results from the infrared behavior
or  from a GYU. (For the
example above, the infrared value is about $195$ GeV.)
 On contrary, if $M_t$ is smaller than
that value, the GYU might be realistic.
Detailed studies on this problem will be
published elsewhere \cite{kubo3}.

\end{document}